\documentclass[prl,amsmath,aps,twocolumn,superscriptaddress]{revtex4}
\usepackage{amsfonts}
\usepackage{graphicx,multirow,color}
\newcommand{\ket}[1]{|#1\rangle}
\newcommand{\bra}[1]{\langle #1|}

\begin{document}
\title{Test of Genuine Multipartite Nonlocality Without Inequality}
\author{Qing Chen}
\affiliation{Centre for quantum technologies, National University of Singapore, 3 Science Drive 2, Singapore 117543, Singapore}
\author{Sixia Yu}
\affiliation{Centre for quantum technologies, National University of Singapore, 3 Science Drive 2, Singapore 117543, Singapore}
\affiliation{
Hefei National Laboratory for Physical Sciences at Microscale and Department of Modern Physics,  University of Science and Technology of China, Hefei, Anhui 230026, China}
\author{ Chengjie Zhang}
\affiliation{Centre for quantum technologies, National University of Singapore, 3 Science Drive 2, Singapore 117543, Singapore}
\author{C.H. Lai}
\affiliation{Centre for quantum technologies, National University of Singapore, 3 Science Drive 2, Singapore 117543, Singapore}
\affiliation{Physics department, National University of Singapore, 3 Science Drive 2, Singapore 117543, Singapore}
\author{C.H. Oh}
\affiliation{Centre for quantum technologies, National University of Singapore, 3 Science Drive 2, Singapore 117543, Singapore}
\affiliation{Physics department, National University of Singapore, 3 Science Drive 2, Singapore 117543, Singapore}

\begin{abstract}
In this letter we propose a set of conditions on the joint probabilities as a test of genuine multipartite nonlocality without inequality. Our test is failed by all non-signaling local models in which even nonlocal correlations among some observables (not all) are allowed as long as these correlations respect the non-signaling principle. A pass of our test by a state therefore indicates that this state cannot be simulated by any non-signaling local models, i.e., the state exhibits genuine multipartite nonlocality. It turns out that all entangled symmetric $n$-qubit ($n\geq 3$)  states pass our test and  therefore are $n$-way nonlocal.
Also we  construct two Bell-type inequalities from our proposed test whose violations indicate genuine multipartite nonlocal correlations.
\end{abstract}

\date{\today}

\maketitle

{\it Introduction.---}
Local measurements performed on composite quantum system can lead to
correlations incompatible with local hidden variable theory \cite{bell}.
This phenomenon is known as quantum nonlocality, which has been recognized as an essential resource for quantum information tasks \cite{Brunner},
such as
quantum key distribution \cite{QKD}, communication complexity \cite{Buhrman},
and randomness generation \cite{Gen}.
Quantum nonlocality can be revealed via several ingenious ways such as the violations of various
Bell inequalities \cite{bell,CHSH}, GHZ paradoxes \cite{GHZ1,GHZ2}, a kind of all-versus-nothing tests,
and Hardy's test of nonlocality without inequality \cite{hardy,hardyn}, a kind of all-versus-something tests.
Most of these tests are designed to rule out `standard' local realistic models, i.e., in the case of multipartite system each observer cannot have nonlocal correlations with any other distant observers.

However, similar to the case of quantum entanglement,
quantum nonlocality displays a much richer and more complex structure
for the multipartite case than the bipartite case.
In the case of three or more observers it is possible to have a hybrid local/nonlocal model in which some observers may share some nonlocal correlations. Still there are quantum correlations that cannot be explained by these more general local models and therefore these states exhibit genuine multipartite or $n$-way nonlocality \cite{Svetlichny}.
Genuine multipartite nonlocality,
being the strongest form of multipartite nonlocality
in which nonlocal correlations are established among all the parties of the system,
naturally attracts much interest recently.
The detection of genuine multipartite  nonlocality also witnesses
genuine multipartite entanglement in a device-independent way.

Svetlichny \cite{Svetlichny} introduced the notion of genuine multipartite nonlocality for the first time and provided a Bell-type inequality to detect genuine tripartite nonlocality.
Recently the result was generalized to arbitrary partite cases \cite{SvetG1} and arbitrary dimensions \cite{SvetG2}.
However Svetlichny's notion of genuine nonlocality is so general that correlations capable of two-way signaling
are allowed among some parties.
As a result, grandfather-type paradoxes arise \cite{Barrett}
and inconsistency from an operational viewpoint appears \cite{Gallego}.
Fortunately, two alternative definitions of genuine multipartite nonlocality can avoid such problems.
One is based on time-ordering \cite{Barrett, Gallego}, in which correlations are non-signaling for at least one observer in each nonlocal group.
The other is based on non-signaling principle \cite{Barrett, Gallego,almeida,sc}, in which the correlations are non-signaling for all observers.
Since the latter way is simpler and rather natural for studying quantum correlations, which are inherently non-signaling, we shall assume that  correlations are non-signaling for all observers.

With the help of this notion of genuine multipartite nonlocality,
completely connected graph states are shown to be genuine multipartite nonlocal \cite{almeida}.
Very recently a modified Hardy-type argument is proposed \cite{Rahaman} to detect genuine multipartite nonlocality, 
however only a restricted set of states satisfies their conditions.
In this letter we give a different set of conditions to detect genuine multipartite nonlocality without inequality.
Those quantum states that pass our test
exhibit genuine multipartite nonlocality.
Out test reduces to the original Hardy's paradox \cite{hardy} for two-particle case and is
different from its various generalizations to multipartite cases in Ref. \cite{hardyn,Rahaman}.
Using our test we prove that all pure entangled symmetric $n$-qubit ($n>2$) states
are genuine multipartite nonlocal. Moreover based on our test
we derive two Bell-type inequalities to detect genuine multipartite nonlocality.

{\it Genuine multipartite nonlocality.--- } Consider a system composed of $n$ spacelike separated subsystems that are labeled with the index set $I=\{1,2,\ldots,n\}$.
Denote by $M_k$ and $r_k$ the measurement setting and the outcome, respectively,
 of the $k$-th part of the system with $k\in I$. The observed correlation is described by the joint probability distribution $P(r_I|M_I)$,
where $r_I=(r_1,\ldots,r_n)$ and $M_I=(M_1,\ldots,M_n)$.
In a standard local realistic model nonlocal correlations are not allowed for any group of observers and the observed joint probability assumes the following form
\begin{equation}
\label{standard}
P(r_I|M_I) = \int \varrho_\lambda \prod_{k=1}^n P_k(r_k|M_k,\lambda) d\lambda,
\end{equation}
where $P_k(r_k|M_k,\lambda)$ is the probability of observer $k$ measuring observable $M_k$ with outcome $r_k$ for given hidden variable $\lambda$ distributed according to $\varrho_\lambda$ with normalization $\int \varrho_\lambda d\lambda =1$.
In the case of three or more observers other possibilities of local models may arise, e.g., two or more observers may share some nonlocal correlations. A genuine multipartite correlation should also exclude such local/nonlocal models.
The most general hybrid local/nonlocal model reads
\begin{equation}
\label{GNNL}
P(r_I|M_I)=\sum_{\alpha \neq \emptyset, \alpha \subset I}
\int \varrho_{\alpha,\lambda} P_{\alpha}(r_{\alpha}|M_{\alpha},\lambda) P_{\bar \alpha}(r_{\bar\alpha}|M_{\bar\alpha},\lambda)d\lambda,
\end{equation}
where we have denoted for every nonempty proper subset $\alpha=\{i_1,\ldots, i_m\}$  of $I$ a restricted outcome vector $r_{\alpha}= (r_{i_1}, \ldots, r_{i_m})$ and a restricted setting vector $M_{\alpha}= (M_{i_1}, \ldots, M_{i_m})$ and by  $P_\alpha(r_\alpha|M_\alpha,\lambda)$ the joint probability of observer $k$ measuring observable $M_k$ with outcome $r_k$ for all observers $k\in \alpha$ for a given hidden variable $\lambda$ distributed according to $\varrho_{\alpha,\lambda}$. We note that $\bar\alpha=I\setminus\alpha$ is also a nonempty proper subset of $I$.
Such correlation may not admit a standard local model and can admit a hybrid local/nonlocal model, i.e.,
nonlocal correlations may be shared among all the observers in group $\alpha$ or $\bar\alpha$, respectively, while the intergroup correlations can only be local.

Furthermore we require that all possible nonlocal correlations in
$P_{\alpha}(r_{\alpha}|M_{\alpha},\lambda)$ and $P_{\bar \alpha}(r_{\bar\alpha}|M_{\bar\alpha},\lambda)$ for any bipartite cut
$\alpha\cup\bar\alpha$ to be non-signaling, i.e., the marginal probability distribution of any part of the
system is independent of the inputs on the remaining part, e.g.,
\begin{eqnarray}
P_{\beta \setminus k}(r_{\beta \setminus k} |M_{\beta \setminus k}, \lambda)=
\sum_{r_k} P_{\beta}(r_{\beta \setminus k}\;r_k |M_{\beta \setminus k}\,M_k, \lambda)  \nonumber \\
\label{NoSignal}
= \sum_{r_k} P_{\beta}(r_{\beta \setminus k}\;r_k |M_{\beta \setminus k}\,M'_k, \lambda)
\end{eqnarray}
for all $k \in \beta$, and $\beta$ can be either $\alpha$ or $\bar \alpha$ with two or more elements.
In order to define genuine multipartite nonlocality in the non-signaling scenario \cite{Barrett, Gallego,almeida,sc},
we should assume the all the bi-local probability distributions are non-signaling.

\textbf{Definition} A joint probability distribution $P(r_I|M_I)$ is called as genuine multipartite nonlocal, or $n$-way nonlocal,
if it  cannot be written as the form  Eq.(\ref{GNNL}) with all possible nonlocal correlations being non-signaling.

{\it Test of genuine multipartite nonlocality without inequality.---}
Consider an $n$-partite system  labeled with $I$ and suppose that
each observer, e.g., the $k$-th local observer, measures two alternative observables, e.g., $\{a_k, b_k\}$, with two outcomes labeled with $\{0, 1\}$. For any $k,k^\prime\in I$ we denote $\bar k=I \setminus\{k\}$ and $\overline{k k'}=I\setminus \{k, k'\}$.
Our start point is the following set of $2n$ joint probabilistic conditions
\begin{subequations}
\label{ours}
\begin{eqnarray}
\label{aaa}
&&P(0_I|a_I)>0, \\
 \label{aab}
&&P(0_I|b_k a_{\bar k})=0,\quad \forall k \in I,  \\
\label{barabb}
&&P \left( 1_{k'} 1_{k}0_{\overline{kk'}}|
  b_{k'} b_{k}a_{\overline{kk'}} \right)=0, \,\, \forall k \in I \setminus \{k' \},
\end{eqnarray}
\end{subequations}
Here $k^\prime\in I$ is fixed and Eq.(\ref{aab}) and Eq.(\ref{barabb}) contain $n$ and $n-1$ conditions respectively.
We shall show that all non-signaling hybrid local/nonlocal models will fail the test, i.e., those $2n$ conditions cannot be satisfied simultaneously by those correlations of form Eq.(\ref{GNNL}).

\textbf{Proposition 1} Any probability distribution that satisfies Eq.(\ref{ours}) is genuine multipartite nonlocal.

We prove this proposition via {\it reductio ad absurdum}.
Suppose a probability distribution satisfies Eq.(\ref{ours}) but is not genuine multipartite nonlocal, i.e.,
it has the form of the right hand side of Eq.(\ref{GNNL}).
Firstly from Eq.(\ref{aaa}) there must exist
some $\alpha_0$ and $\lambda_0$ making
$P_{\alpha_0}(0_{\alpha_0}|a_{\alpha_0},\lambda_0)>0$ and $P_{\bar\alpha_0}(0_{\bar \alpha_0}|a_{\bar\alpha_0},\lambda_0)>0$.
Combining this condition with Eq.(\ref{aab}) we obtain
\begin{eqnarray}
\label{kkk}
P_{\alpha_0}(0_k 0_{\alpha_0 \setminus k }|b_k a_{\alpha_0\setminus k }, \lambda_0)=0, && \forall k \in \alpha_0 \\
\label{barkkk}
P_{\bar\alpha_0}(0_k 0_{\bar \alpha_0 \setminus k}|b_k a_{\bar \alpha_0 \setminus k }, \lambda_0)=0, && \forall k \in \bar \alpha_0.
\end{eqnarray}
Since both $\alpha_0$ and  $\bar\alpha_0$ are nonempty,
without loss of generality, we suppose  $k'\in \bar\alpha_0$ and let $j\in \alpha_0$ be an arbitrary element in $\alpha_0$.
From Eq.(\ref{barabb}) we obtain either
$P_{\bar\alpha_0}(1_{k'} 0_{\bar \alpha_0 \setminus k'}|b_{k'} a_{\bar \alpha_0 \setminus k' }, \lambda_0)=0$ or
$P_{\alpha_0}(1_j 0_{\alpha_0 \setminus j}|b_j a_{\alpha_0 \setminus j }, \lambda_0)=0$.
If the latter equation holds,
by combining Eq.(\ref{kkk}) and using the non-signaling condition Eq.(\ref{NoSignal}),
we obtain
$P_{\alpha_0 \setminus j}( 0_{\alpha_0 \setminus j}|a_{\alpha_0 \setminus j }, \lambda_0)=
P_{\alpha_0}(1_j 0_{\alpha_0 \setminus j}|b_j a_{\alpha_0 \setminus j }, \lambda_0) +
P_{\alpha_0}(0_j 0_{\alpha_0 \setminus j}|b_j a_{\alpha_0 \setminus j }, \lambda_0)
=0$.
However due to non-signaling condition again,
$P_{\alpha_0\setminus j}( 0_{\alpha_0 \setminus j}|a_{\alpha_0 \setminus j }, \lambda_0)\geq
P_{\alpha_0}(0_{\alpha_0}|a_{\alpha_0 }, \lambda_0) >0$,
thus we find a contradiction.
If the former equation holds, by combining Eq.(\ref{barkkk}) a similar contradiction can also be made.

Interestingly for two-particle case the test Eq.(\ref{ours}) reduces to
original Hardy's paradox \cite{hardy}.
For the case of $n>2$, by replacing the set of conditions in Eq.(\ref{barabb}) with
a single condition $P \left( 1_I|  b_I \right)=0$, our test
Eq.(\ref{ours}) reduces to the `standard' Hardy's nonlocality conditions \cite{hardyn} for multipartite case,
which can only detect `standard' multipartite nonlocality.
Thus our test can be regarded as a natural generalization of Hardy's paradox to detect genuine multipartite nonlocality.

Our test is failed by all non-signaling local models but there are quantum states that pass it.
For quantum systems, only genuine entangled states may exhibit genuine multipartite nonlocality. For an arbitrary $n$-partite state $\rho$, to test its $n$-way nonlocality, one must find two measurement settings $\{\ket{a_i}, \ket{b_i}\}$ for each particle $i$ satisfying
\begin{eqnarray}
&&\bra{a_I} \rho \ket{a_I} >0,  \nonumber \\
 \label{quanaaa}
&& \bra{b_k a_{\bar k}}  \rho   |b_k a_{\bar k} \rangle= 0, \quad \forall k \in I,  \\
&& \bra{\bar b_1 \bar b_{k}a_{\overline{1k}}}   \rho |\bar b_1 \bar b_{k}a_{\overline{1k}} \rangle= 0, \quad \forall k \in \bar 1, \nonumber
\end{eqnarray}
where $|a_\alpha\rangle=\otimes_{k\in \alpha}|a_k\rangle$ and $|\bar b_k\rangle$ being  orthogonal to $|b_k\rangle$.
Note that for simplicity we just let $k'=1$.
There exists a systematic way to construct such
quantum states by choosing pairs of non-commuting observables, i.e.,
$[A_i, B_i]\neq 0$, for each particle $i$.
Let $\{\ket{a_i}, \ket{\bar a_i}\}$ and
$\{\ket{b_i}, \ket{\bar b_i}\}$ be two eigenstates for $A_i$ and $B_i$ with outcomes $\{+1, -1\}$ respectively.
We denote by $\mathbb{P}$
the subspace  of the Hilbert space of the system spanned by $2n$ linearly independent vectors
$\{\ket{a_I}, \ket{b_1 a_{\bar 1}}, \ldots,  \ket{b_n a_{\bar n}},
\ket{\bar b_1 \bar b_{2}a_{\overline{12}}},\ldots, \ket{\bar b_1 \bar b_{n}a_{\overline{1n}}}  \}$
and denote by $\mathbb{I_P}$ the projection to this subspace.
It is easy to see that ${\mathbb{P}}$ contains one and only one pure state $\ket{\phi}$ satisfying Eq.(\ref{quanaaa}).
In the case of two qubits \cite{hardy2},  the dimension of the subspace ${\mathbb{P}}$ is equal to that of the system,
which implies that there is a unique $\ket{\phi}$ satisfying Eq.(\ref{quanaaa}).
However, in the case of $n \geq 3$, the dimension of ${\mathbb{P}}$ being smaller than that of the system,
any quantum state $\rho$ (pure or mixed) satisfying $\mathbb{I_P}\,\rho\,\mathbb{I_P} \propto \ket{\phi}\bra{\phi}$ can also satisfy Eq.(\ref{quanaaa}).
A more important issue is, however, the converse problem, i.e.,
for a given state, how to
ascertain the corresponding measurement settings such that Eq.(\ref{quanaaa}) are satisfied.

{\it Permutation symmetric states.--- }
A pure symmetric $n$-qubit state can be written as
\begin{equation}
\ket{\psi}= \sum_{k=0}^n h_k\sum_{\alpha\subseteq I,|\alpha|=k} |0_{\bar\alpha} 1_{\alpha}\rangle.
\end{equation}
First we note that the closest product state of a symmetric state, whose inner product with $|\psi\rangle$ is the largest among all possible product states, is also a symmetric state \cite{cps}.
Thus without loss of generality, we suppose that $\ket{0_I}$ is already the closest product state of $|\psi\rangle$. The computational basis determined by the closest product state is a magic basis \cite{Yu} in which $h_0\not=0$ and $h_1=0$.
All entangled symmetric states have been proved to exhibit `standard' nonlocality \cite{sym}.
Here we have

\textbf{Proposition 2} All pure multipartite entangled permutation symmetric $n$-qubit ($n\geq 3$) states
are genuine multipartite nonlocal.

We prove this proposition by showing that all pure entangled symmetric states pass
our test by choosing the measurement settings properly.
We choose the measurement settings on all the qubits $k\in \bar 1$ to be the same,
especially,
let
$\ket{a_1}=\ket{0}+x_1^* \ket{1}$, $\ket{b_1}=\ket{0}+y_1^* \ket{1}$,
 and
$\ket{a_i}=\ket{a}=\ket{0}+x^* \ket{1}$,
$\ket{b_i}=\ket{0}+y^* \ket{1}$ for $ 2 \leq i \leq n$,
where $x, y, x_1, y_1$ are complex numbers and $x$ is finite.
Then Eq.(\ref{quanaaa}) reduces to the following four conditions as
\begin{subequations}
\label{symm}
\begin{eqnarray}
\label{2a1a2}
&& \bra{a_1a_2}\psi_{12}\rangle \neq 0, \\
\label{2b1b2}
&& \bra{\bar b_1 \bar b_2}\psi_{12}\rangle = \bra{a_1b_2}\psi_{12}\rangle=\bra{b_1a_2}\psi_{12}\rangle =0,
\end{eqnarray}
\end{subequations}
where $J=I\setminus \{1,2\}$,
$$\ket{\psi_{12}}=\bra{a_J}\psi\rangle=c_0\ket{00}+c_1(\ket{01}+\ket{10})+c_2\ket{11}$$
is an unnormalized symmetric state (or zero),
and $$c_i=\sum_{k=0}^{n-2} h_{k+i} x^k C_{n-2}^k,\quad 0 \leq i \leq 2.$$
Note that Eq.(\ref{symm}) has the similar form to the original Hardy's paradox,
however it should be kept in mind here
that both $\ket{\psi_{12}}$ and $\ket{a_2}$ depend on $x$.
Obviously, to satisfy Eq.(\ref{symm}) the 2-particle state $\ket{\psi_{12}}$  cannot be a product state or zero. These two cases can be avoided by the following observation.

\textbf{Observation} For any genuine entangled symmetric pure $n$-qubit state $\ket{\psi}$,
there only exist finite local projections $\ket{a}$ making the projected state
$\ket{\psi_{12}}=\bra{a_J}\psi\rangle$ a product state or zero.

The necessary and sufficient condition for $\ket{\psi_{12}}$ being a product state or zero
is $c_1^2-c_0c_2=0$,
which is an algebraic equation for $x$ of a degree at most $2(n-2)$.
In general there are only finite solutions of $x$ (corresponding to finite local projections) except for the case that $c_1^2-c_0c_2$ vanishes identically, i.e.,
all its coefficients of $x^m$ are zero
\begin{equation}
\sum_{k+k'=m} (h_{k+1} h_{k'+1}- h_{k} h_{k'+2}) C_{n-2}^k C_{n-2}^{k'}=0
\end{equation}
for all $0 \leq m \leq 2(n-2)$.
This exceptional case allows $x$ to take an arbitrary value.
However, this can only happen when $\ket{\psi}$ is a product state for the following
reasons.
Starting from the case $m=0$ with $k=k^\prime=0$, by noting that $h_0 \neq 0$ and $h_1=0$, one gets $h_2=0$.
Recursively, with the increasing of $m$ by $1$,
from $h_0 \neq 0$ and $h_j=0$ for all $j \leq m+1$, one obtains $h_{m+2}=0$.
As a result $h_i=0$ for $i>0$ and $\ket{\psi}$ can only be a product state.

In what follows we shall show that by choosing any complex $x$ with a fixed phase $e^{iw}=x/|x|$ such that ${h_0^*h_2}e^{-iw}$ is not real and excluding a finite number of values of $|x|$,
any pure multipartite entangled symmetric state will  pass our nonlocality test.
From   Eq.(\ref{2b1b2}) we find the solution of $y_1,y,x_1$ as an (implicit) function of $x$ respectively as
$$y_1=-\frac{c_0+xc_1}{c_1+xc_2}, \,\, y=\frac{c_2^*-y_1c_1^*}{c_1^*-y_1c_0^*}, \,\,
x_1=-\frac{c_0+yc_1}{c_1+yc_2}. $$
By substituting them into  $\bra{a_1a_2}\psi_{12}\rangle$, the condition $\bra{a_1a_2}\psi_{12}\rangle=0$ leads to
$F(x,x^*)(c_0c_2-c_1^2) =0$
where
$$F(x,x^*)= c_1c_2^*+c_0c_1^*+(|c_2|^2-|c_0|^2)x-(c_1^*c_2+c_0^*c_1)x^2$$
is a polynomial of $x$ and $x^*$ with its constant term vanishing since $h_1=0$. The coefficients of the linear term $x$ and $x^*$ read $(n-1)|h_2|^2-|h_0|^2$ and $h_0h_2^*$ respectively. As a result if $h_2=0$ then $F(x,x^*)$ is not identical to zero and for any fixed $w$
there are only a finite number of solutions to $F(x,x^*)=F_w(|x|)=0$ for $|x|$. If $h_2\neq 0$ then as long as $h_0h_2^*e^{-i2w}$ is not real, i.e., $x^2$ has a phase that is different from that of $h_0^*h_2$, then $F(x,x^*)$ is not identically vanishing and there are only a finite number of solutions to $F(x,x^*)=F_w(|x|)=0$ for $|x|$.

As a result there exist an infinite number of $x$ that do not satisfy $F(x,x^*)=0$, combining with the Observation, i.e.,
there only exist finite solutions for $c_0c_2-c_1^2=0$,
finally we get the conclusion that there exist infinite measurement settings
satisfying Eq.(\ref{symm}) for any pure multipartite entangled symmetric state.

We note that our result above strengthens previous works \cite{sym,Yu}
in two folds. On the one hand all pure multipartite entangled symmetric
 states are proved to exhibit genuine multipartite nonlocality and on the other hand,
this nonlocality can be revealed by a test without inequality.

In practice, to detect $n$-way nonlocality of a given symmetric state, one can solve the equations
Eq.(\ref{symm}) directly under its computational basis. As demonstrations, we shall present two examples, namely the generalized GHZ state and the $W$ state.
For a generalized GHZ state
$\ket{G_n(\theta)}= \cos \theta \ket{0_I} + \sin \theta \ket{1_I}$,
where $0 < \theta < \pi/2$,
one obtains $c_0=\cos\theta$, $c_1=0$ and $c_2=\sin\theta x^{n-2}$.
To satisfy Eq.(\ref{2b1b2}),
one gets
$y_1=-\cot \theta/x^{n-1}$,
$y= x^{n-1} x^{*(n-2)} \tan^2 \theta $,
and
$x_1^{-1}=-\tan^3 \theta|x|^{2n-4}x^{n-1}$.
Thus
$$|\bra{a_I}G_n(\theta)\rangle|^2 =
\frac{\cos^2\theta(1 -  \cot^2 \theta/|x|^{2n-4})^2}{(1+|x_1|^2)(1+|x|^2)^{n-1}},$$
which is always larger than zero except for the cases $|x|=  0,  \cot^{1/(n-2)}\theta , \infty$.
For $W$ state
$|W_n\rangle = \frac{1}{\sqrt{n}}\sum_{k=1}^n|0_{\bar k} 1_{k}\rangle,$
one obtains
$c_0=x(n-2)/\sqrt{n}$, $c_1=1/\sqrt{n}$, $c_2=0$.
From Eq.(\ref{2b1b2}),
we have
$y_1=-x(n-1)$,
$y= \frac{x(n-1)}{1+(n-1)(n-2)|x|^2} $,
and
$x_1=-x(n-2)-y$.
Thus
\begin{equation}
|\bra{a_I} W_n\rangle|^2 =\frac{|x-y|^2}{n(1+|x_1|^2)(1+|x|^2)^{n-1}},\nonumber
\end{equation}
which equals to zero only for the cases $|x|= 0,  \sqrt{ \frac{1}{n-1}}, \infty$.

{\it The Bell-type inequalities to detect genuine multipartite nonlocality.---}
Similar to the relation between Hardy's paradox and Hardy's inequality,
one can immediately obtain a Bell-type inequality from  Eq.(\ref{ours}) as follows
(see Appendix for proof)
\begin{eqnarray}
\label{ineq1}
&&P(0_I|a_I)-\sum_{k \in I} P(0_I|b_k a_{\bar k})- \nonumber \\
&&\sum_{k \in I\setminus \{k'\}} P \left( 1_{k'} 1_{k}0_{\overline{kk'}}|  b_{k'} b_{k}a_{\overline{kk'}} \right)\le 0.
\end{eqnarray}
Note that this inequality is stronger than the test Eq.(\ref{ours})
for detecting genuine multipartite nonlocality, i.e.,
any quantum state satisfies Eq.(\ref{ours}) should violate this inequality, however the converse may not be true.

More interestingly,
by symmetrizing the last term of the Bell-type inequality Eq.(\ref{ineq1}) for all particles,
we obtain
a different Bell-type inequality (see Appendix for proof) to detect genuine multipartite nonlocality as follows,
\begin{eqnarray}\label{ineq2}
&&P(0_I|a_I)-\sum_{k \in I} P(0_I|b_k a_{\bar k})- \nonumber \\
&&\frac{1}{n-1}\sum_{k,k' \in I, k\neq k'} P \left( 1_{k'} 1_{k}0_{\overline{kk'}}|  b_{k'} b_{k}a_{\overline{kk'}} \right) \leq 0.
\end{eqnarray}
Note that in the non-signaling scenario and in the case of $n=3$,
this inequality is equivalent to the one found in Ref \cite{Barrett},
which is found numerically to be violated by all pure tripartite entangled states.

{\it Discussions.---}
All pure entangled  states exhibit standard nonlocality \cite{pr, Yu}
and it is an open problem whether all pure genuine multipartite entangled states
are genuine multipartite nonlocal \cite{Brunner}.
We have made a progress towards solving this problem by proposing
a test without inequality, with the help of which we have shown
that all entangled symmetric states are genuine multipartite nonlocal.
In the case of asymmetric states,
we have tried 50000 randomly chosen pure genuine multipartite entangled states
for three and four qubits each to our test Eq.(\ref{ours}) and all states turn out to be genuine multipartite nonlocal.
Thus we conjecture that
all pure genuine multipartite entangled states are genuine multipartite nonlocal in the non-signaling scenario
and can be detected by our test without inequality.
However, the analytic proof remains a challenging issue.

This work is supported by National Research Foundation
and Ministry of Education, Singapore (Grant No.
WBS: R-710-000-008-271) and NSF of China (Grant No.
11075227).

\setcounter{equation}{0}
\renewcommand{\theequation}{S\arabic{equation}}
\begin{widetext}

{\it Appendix.---} Here we shall prove that our Bell-type inequalities Eq.(\ref{ineq1}) and Eq.(\ref{ineq2})
are satisfied by all non-signaling local models.
Recall that we label $n$ qubits with the index set $I=\{1,2,\ldots,n\}$ and for a nonempty subset $\alpha\subset I$ we denote $\bar\alpha=I\setminus \alpha$ and $|\alpha|$ as its number of elements.
By linearity we only have to prove that
the correlation
$P_{\alpha}^{\lambda}(r_I|M_I) = P_{\alpha}(r_{\alpha}|M_{\alpha},\lambda) P_{\bar \alpha}(r_{\bar\alpha}|M_{\bar\alpha},\lambda)$
satisfies the inequalities for any given $\alpha$ and $\lambda$.

{\it The proof of Eq.(\ref{ineq1}).---}
Note that $k'$ can be an arbitrary number with $1 \leq k' \leq n$.
However since both $\alpha$ and  $\bar\alpha$ are nonempty subsets of $I$,
without loss of generality, we
suppose  $k'\in \bar\alpha$ and thus $\alpha$ contains at least one element, denoted as $j$.
For the probability distribution $P_{\alpha}^{\lambda}(r_I|M_I)$,
the left hand of Eq.(\ref{ineq1}) becomes

\begin{eqnarray}
&&P_{\alpha}(0_{\alpha}|a_{\alpha},\lambda)P_{\bar \alpha}(0_{\bar\alpha}|a_{\bar\alpha},\lambda)-
\sum_{k \in \alpha}      P_{ \alpha}(0_k 0_{\alpha \setminus k} |b_k a_{\alpha \setminus k},\lambda)
P_{\bar \alpha}(0_{\bar\alpha}|a_{\bar\alpha},\lambda)-
\sum_{k \in \bar \alpha}
P_{ \alpha}(0_{\alpha}|a_{\alpha},\lambda)P_{\bar \alpha}(0_k 0_{\bar \alpha \setminus k}|b_k a_{\bar \alpha \setminus k }, \lambda) \nonumber \\
&&
\label{pp1}
-\sum_{k \in \alpha}  P_{\alpha}(1_k 0_{\alpha \setminus k} |b_k a_{\alpha \setminus k},\lambda)
P_{\bar \alpha}(1_{k'} 0_{\bar\alpha \setminus k'} |b_{k'} a_{\bar\alpha \setminus k'},\lambda)-
\sum_{k \in \bar \alpha} P_{ \alpha}(0_{\alpha}|a_{\alpha},\lambda)
P_{\bar \alpha}(1_{k'}1_k 0_{\bar\alpha \setminus \{k,k'\}} |b_{k'} b_k a_{\bar\alpha \setminus \{k,k'\}},\lambda) \\
&\leq&
 \label{S0}
P_{\alpha}(0_{\alpha}|a_{\alpha},\lambda)P_{\bar \alpha}(0_{\bar\alpha}|a_{\bar\alpha},\lambda)-
  P_{ \alpha}(0_j 0_{\alpha \setminus j} |b_j a_{\alpha \setminus j},\lambda)P_{\bar \alpha}(0_{\bar\alpha}|a_{\bar\alpha},\lambda)-
 P_{\alpha}(0_{\alpha}|a_{\alpha},\lambda)P_{\bar \alpha}(0_{k'} 0_{\bar \alpha \setminus k'}|b_{k'} a_{\bar \alpha \setminus k' }, \lambda)
 \\
&&- P_{\alpha}(1_j 0_{\alpha \setminus j} |b_j a_{\alpha \setminus j},\lambda)
P_{\bar \alpha}(1_{k'} 0_{\bar\alpha \setminus k'} |b_{k'} a_{\bar\alpha \setminus k'},\lambda) \nonumber \\
&\leq&
 \label{S1}
\Big( P_{ \alpha}(0_{\alpha}|a_{\alpha},\lambda)-
  P_{\alpha}(0_j 0_{\alpha \setminus j} |b_j a_{\alpha \setminus j},\lambda)-
 \min \big[ P_{ \alpha}(0_{\alpha}|a_{\alpha},\lambda),P_{ \alpha}(1_j 0_{\alpha \setminus j} |b_j a_{\alpha \setminus j},\lambda) \big] \Big)
 P_{\bar \alpha}(0_{\bar \alpha}|a_{\bar \alpha }, \lambda)  \\
&=&
\label{S2}
\Bigg\{ \begin{array}{ll}
- P_{ \alpha}(0_j 0_{\alpha \setminus j} |b_j a_{\alpha \setminus j},\lambda)P_{\bar \alpha}(0_{\bar\alpha}|a_{\bar\alpha},\lambda) \leq 0 &
\textrm{if $ P_{ \alpha}(0_{\alpha}|a_{\alpha},\lambda)\leq P_{ \alpha}(1_j 0_{\alpha \setminus j} |b_j a_{\alpha \setminus j},\lambda)$.}\\
\Big(P_{ \alpha}(0_{\alpha}|a_{\alpha},\lambda)-
P_{ \alpha \setminus j}( 0_{\alpha \setminus j} | a_{\alpha \setminus j},\lambda) \Big)
P_{\bar \alpha}(0_{\bar\alpha}|a_{\bar\alpha},\lambda) \leq 0 &
\textrm{if $ P_{ \alpha}(0_{\alpha}|a_{\alpha},\lambda)> P_{ \alpha}(1_j 0_{\alpha \setminus j} |b_j a_{\alpha \setminus j},\lambda)$.}
\end{array}
\end{eqnarray}
Note that to get Eq. (\ref{S0}), we firstly drop the last term of Eq. (\ref{pp1}),
and then for the second, third, and fourth terms in Eq. (\ref{pp1}),
we just keep the item of $k=j$, $k=k'$, and $k=j$ respectively and drop all other items.
To obtain Eq. (\ref{S1}), we use the fact that
$P_{\bar \alpha}(0_{k'} 0_{\bar \alpha \setminus k'}|b_{k'} a_{\bar \alpha \setminus k' }, \lambda)+
P_{\bar \alpha}(1_{k'} 0_{\bar\alpha \setminus k'} |b_{k'} a_{\bar\alpha \setminus k'},\lambda)
\geq  P_{\bar \alpha}(0_{\bar \alpha}|a_{\bar \alpha }, \lambda)$, which is deduced by non-signaling principle.

{\it The proof of Eq.(\ref{ineq2}).---}
For the distribution $P_{\alpha}^{\lambda}(r_I|M_I)$,
the left hand of Eq.(\ref{ineq2}) becomes
\begin{eqnarray}
&& P_{ \alpha}(0_{\alpha}|a_{\alpha},\lambda)P_{\bar \alpha}(0_{\bar\alpha}|a_{\bar\alpha},\lambda)-
\sum_{k \in \alpha}      P_{ \alpha}(0_k 0_{\alpha \setminus k} |b_k a_{\alpha \setminus k},\lambda)
P_{\bar \alpha}(0_{\bar\alpha}|a_{\bar\alpha},\lambda)-
\sum_{k \in \bar \alpha} P_{\alpha}(0_{\alpha}|a_{\alpha},\lambda)
P_{\bar \alpha}(0_k 0_{\bar \alpha \setminus k}|b_k a_{\bar \alpha \setminus k }, \lambda) \nonumber \\
&&
\label{pp2}
-\frac{1}{n-1} \Bigg\{ \sum_{k \in \alpha, k' \in \bar \alpha}  P_{ \alpha}(1_k 0_{\alpha \setminus k} |b_k a_{\alpha \setminus k},\lambda)
P_{\bar \alpha}(1_{k'} 0_{\bar\alpha \setminus k'} |b_{k'} a_{\bar\alpha \setminus k'},\lambda)
+ \\
&&
\sum_{k,k' \in \bar \alpha} P_{\alpha}(0_{\alpha}|a_{\alpha},\lambda)
P_{\bar \alpha}(1_{k}1_{k'} 0_{\bar\alpha \setminus \{k,k'\}} |b_{k} b_{k'} a_{\bar\alpha \setminus \{k,k'\}},\lambda)
+\sum_{k,k' \in \alpha}
P_{ \alpha}(1_{k}1_{k'} 0_{\alpha \setminus \{k,k'\}} |b_{k} b_{k'} a_{\alpha \setminus \{k,k'\}},\lambda)
P_{\bar \alpha}(0_{\bar\alpha}|a_{\bar\alpha},\lambda)\Bigg\}
\nonumber \\
&\leq&
P_{ \alpha}(0_{\alpha}|a_{\alpha},\lambda)P_{\bar \alpha}(0_{\bar\alpha}|a_{\bar\alpha},\lambda)-
P_{ \alpha}(0_{\alpha}|a_{\alpha},\lambda)\sum_{k' \in \bar \alpha}
P_{\bar \alpha}(0_{k'} 0_{\bar \alpha \setminus k'}|b_{k'} a_{\bar \alpha \setminus k' }, \lambda)-
\nonumber \\
&&
\label{S20}
\sum_{k \in \alpha} \Big\{  P_{ \alpha}(0_k 0_{\alpha \setminus k} |b_k a_{\alpha \setminus k},\lambda)
P_{\bar \alpha}(0_{\bar\alpha}|a_{\bar\alpha},\lambda)
+  P_{ \alpha}(1_k 0_{\alpha \setminus k} |b_k a_{\alpha \setminus k},\lambda) \sum_{k' \in \bar \alpha} \frac{1}{n-1}
P_{\bar \alpha}(1_{k'} 0_{\bar\alpha \setminus k'} |b_{k'} a_{\bar\alpha \setminus k'},\lambda)
\Big \}
\\
&\leq&
P_{ \alpha}(0_{\alpha}|a_{\alpha},\lambda) \Bigg\{ P_{\bar \alpha}(0_{\bar\alpha}|a_{\bar\alpha},\lambda)-
\sum_{k' \in \bar \alpha}
P_{\bar \alpha}(0_{k'} 0_{\bar \alpha \setminus k'}|b_{k'} a_{\bar \alpha \setminus k' }, \lambda)
\nonumber \\
&&
\label{max2}
-|\alpha|
P_{ \alpha}(0_{\alpha } |a_{\alpha},\lambda) \min \Big[
P_{\bar \alpha}(0_{\bar\alpha}|a_{\bar\alpha},\lambda),
\frac{1}{n-1} \sum_{k' \in \bar \alpha}
P_{\bar \alpha}(1_{k'} 0_{\bar\alpha \setminus k'} |b_{k'} a_{\bar\alpha \setminus k'},\lambda)
\Big ] \Bigg\}.
\end{eqnarray}
To get Eq.(\ref{S20}),
we just drop the last two terms of Eq.(\ref{pp2}).
To obtain Eq.(\ref{max2}), we use the fact that
$P_{ \alpha}(0_{k} 0_{ \alpha \setminus k}|b_{k} a_{ \alpha \setminus k }, \lambda)+
P_{  \alpha}(1_{k} 0_{\alpha \setminus k} |b_{k} a_{\alpha \setminus k},\lambda)
\geq  P_{ \alpha}(0_{ \alpha}|a_{ \alpha }, \lambda)$ for all $k\in \alpha$.
Now if $$ P_{\bar \alpha}(0_{\bar\alpha}|a_{\bar\alpha},\lambda)\leq \frac{1}{n-1}
\sum_{k' \in \bar \alpha}
P_{\bar \alpha}(1_{k'} 0_{\bar\alpha \setminus k'} |b_{k'} a_{\bar\alpha \setminus k'},\lambda)$$
then Eq.(\ref{max2}) becomes
\begin{eqnarray}
P_{ \alpha}(0_{\alpha}|a_{\alpha},\lambda) \Big\{ (1-|\alpha|)P_{\bar \alpha}(0_{\bar\alpha}|a_{\bar\alpha},\lambda)-
\sum_{k' \in \bar \alpha}
P_{\bar \alpha}(0_{k'} 0_{\bar \alpha \setminus k'}|b_{k'} a_{\bar \alpha \setminus k' }, \lambda) \Big\}
\leq 0,
\end{eqnarray}
because $|\alpha|$ is always equal to or larger than 1.
If $$ P_{\bar \alpha}(0_{\bar\alpha}|a_{\bar\alpha},\lambda)> \frac{1}{n-1}
\sum_{k' \in \bar \alpha}
P_{\bar \alpha}(1_{k'} 0_{\bar\alpha \setminus k'} |b_{k'} a_{\bar\alpha \setminus k'},\lambda)$$
then Eq. (\ref{max2}) becomes
\begin{eqnarray}
&&
P_{\alpha}(0_{\alpha}|a_{\alpha},\lambda)
\left\{
P_{\bar\alpha}(0_{\bar\alpha}|a_{\bar\alpha},\lambda)-
\sum_{k' \in \bar \alpha} P_{\bar\alpha}(0_{k'} 0_{\bar \alpha \setminus k'}|b_{k'} a_{\bar \alpha \setminus k' }, \lambda)
-\frac{|\alpha|}{n-1}
\sum_{k' \in \bar \alpha}
P_{\bar\alpha}(1_{k'} 0_{\bar\alpha \setminus k'} |b_{k'} a_{\bar\alpha \setminus k'},\lambda)
\right\} \nonumber \\
&\leq&
P_{\alpha}(0_{\alpha}|a_{\alpha},\lambda)
P_{\bar\alpha}(0_{\bar\alpha}|a_{\bar\alpha},\lambda)\Big(1-\frac{|\alpha||\bar \alpha|}{n-1}\Big) \leq 0,
\end{eqnarray}
in which the first inequality is due to  $|\alpha| \leq n-1$ and
$P_{\bar \alpha}(0_{k'} 0_{\bar \alpha \setminus k'}|b_{k'} a_{\bar \alpha \setminus k' }, \lambda)+
P_{\bar \alpha}(1_{k'} 0_{\bar\alpha \setminus k'} |b_{k'} a_{\bar\alpha \setminus k'},\lambda)
\geq  P_{\bar \alpha}(0_{\bar \alpha}|a_{\bar \alpha }, \lambda)$ for all $k' \in \bar \alpha$ while the second inequality is due to the fact that $|\alpha||\bar \alpha| \geq n-1 $ for any partition of $I$ into two nonempty subsets $\alpha$ and $\bar \alpha$.

\end{widetext}
\end{document}